\documentclass[submission,copyright,creativecommons]{eptcs}
\providecommand{\event}{FICS 2012} 
\usepackage{breakurl}             

\usepackage{savesym}
\usepackage{amsmath,amssymb,amsthm}
\savesymbol{iint}
\usepackage{txfonts}
\restoresymbol{TXF}{iint}

\usepackage[all]{xy}

\newcommand{\squishlist}{
  \begin{list}{$\bullet$}
   { \setlength{\itemsep}{0pt}      \setlength{\parsep}{2pt}
     \setlength{\topsep}{2pt}       \setlength{\partopsep}{0pt}
     \setlength{\leftmargin}{1.2em} \setlength{\labelwidth}{1em}
     \setlength{\labelsep}{0.5em} } }

\newcommand{\squishlisttwo}{
  \begin{list}{$\bullet$}
   { \setlength{\itemsep}{0pt}    \setlength{\parsep}{0pt}
     \setlength{\topsep}{0pt}     \setlength{\partopsep}{0pt}
     \setlength{\leftmargin}{2em} \setlength{\labelwidth}{1.5em}
     \setlength{\labelsep}{0.5em} } }

\newcommand{\squishend}{
   \end{list}  }

\newtheorem{theorem}{Theorem}[section]
\newtheorem{lemma}[theorem]{Lemma}

\newtheorem{proposition}[theorem]{Proposition}

\theoremstyle{definition}
\newtheorem{definition}[theorem]{Definition}
\theoremstyle{remark}

\newtheorem{example}[theorem]{Example}

\newcommand{\coleq}{\mathrel{\mathop:}=}

\newcommand{\Mod}{\hbox{\rm Mod}}
\newcommand{\BMod}{\hbox{\rm BMod}}
\newcommand{\dom}{\hbox{\rm dom}}
\newcommand{\codom}{\hbox{\rm codom}}
\newcommand{\End}{\hbox{\rm End}}
\newcommand{\Mon}{\hbox{\rm Mon}}
\newcommand{\Ar}{\hbox{\rm Ar}}
\newcommand{\WRep}{\hbox{\rm RMon}}

\newcommand{\rar}{\longrightarrow}

\newcommand{\app}{{\mathsf{app}}}
\newcommand{\abs}{{\mathsf{abs}}}

\newcommand{\join}{{\mathsf{join}}}

\newcommand{\Cat}[1]{\mathsf{#1}}
\newcommand{\CC}{\Cat{C}}
\newcommand{\DD}{\Cat{D}}

\newcommand{\Fin}{\Cat{Fin}}
\newcommand{\Set}{{\mathsf{Set}}}
\newcommand{\subst}{{\mathsf{subst}}}

\usepackage{hyperref}
\hypersetup{
    bookmarks=true,         
    unicode=false,          
    pdftoolbar=true,        
    pdfmenubar=true,        
    pdffitwindow=false,     
    pdfstartview={FitH},    
    pdftitle={My title},    
    pdfauthor={Author},     
    pdfsubject={Subject},   
    pdfcreator={Creator},   
    pdfproducer={Producer}, 
    pdfkeywords={keyword1} {key2} {key3}, 
    pdfnewwindow=true,      
    colorlinks=true,       
    linkcolor=blue,          
    citecolor=blue,        
    filecolor=magenta,      
    urlcolor=cyan           
}

\title{Initial Semantics for Strengthened Signatures}
\author{Andr\'e Hirschowitz
\institute{Laboratoire J.-A.~Dieudonn\'e\\
Universit\'e de Nice - Sophia Antipolis\\
France}
\email{ah@unice.fr}
\and Marco Maggesi
\institute{Dipartimento di Matematica ``U.~Dini''\\
Universit\`a degli Studi di Firenze\\
Italy}
\email{maggesi@math.unifi.it}}

\def\titlerunning{Initial Semantics for Strengthened Signatures}
\def\authorrunning{A.~Hirschowitz, M.~Maggesi}

\begin{document}
\maketitle

\begin{abstract}
  We give a new general definition of arity, yielding the companion
  notions of signature and associated syntax. This setting is modular
  in the sense requested by \cite{GU06}: merging two extensions of
  syntax corresponds to building an amalgamated sum. These signatures
  are too general in the sense that we are not able to prove the
  existence of an associated syntax in this general context.  So we
  have to select arities and signatures for which there exists the
  desired initial monad. For this, we follow a track opened by Matthes
  and Uustalu \cite {MU03}: we introduce a notion of strengthened
  arity and prove that the corresponding signatures have initial
  semantics (i.e.~associated syntax).  Our strengthened arities admit
  colimits, which allows the treatment of the $\lambda$-calculus with
  explicit substitution in the spirit of \cite {GU06}.
\end{abstract}

\section{Introduction}


Many programming or logical languages allow constructions which bind
variables and this higher-order feature causes much trouble in the
formulation, the understanding and the formalization of the theory of
these languages. For instance, there is no universally accepted
discipline for such formalizations: that is precisely why the POPLmark
Challenge \cite{PoplMark} offers benchmarks for testing old and new
approaches. Although this problem may ultimately concern typed
languages and their operational semantics, it already concerns untyped
languages.  In this work, we extend to new kinds of constructions our
treatment of higher-order abstract syntax \cite{HM}, based on modules
and linearity.


First of all, we give a new general definition of arity, yielding the
companion notion of signature.  The notion is coined in such a way to
induce a companion notion of representation of an arity (or of a
signature) in a monad: such a representation is a morphism among
modules over the given monad, so that an arity simply assigns two
modules to each monad.  There is a natural category of such
representations of a signature and whenever it exists, the initial
representation deserves the name of syntax associated with the given
signature.  This approach enjoys modularity in the sense introduced by
\cite{GU06}: in our category of representations, merging two
extensions of a syntax corresponds to building an amalgamated sum.

Our notion of arity (or signature) is too general in the sense that we
are not able to build, for each signature, a corresponding initial
representation.  Following a track opened in Matthes-Uustalu
\cite{MU03}, we define a fairly general notion of \emph{strengthened}
arity, yielding the corresponding notion of strengthened signature.
Our main result (Theorem \ref{t:initsynt}) says that any
strengthened signature yields the desired initial representation.  As
usual, this initial object is built as a minimal fixpoint.


Understanding the syntax of the lambda-calculus with explicit
substitution was already done in \cite{GU06}, where the arity for this
construction was identified as a coend, hence a colimit, of elementary
arities (see Section \ref{ss:ex-expl-subst}).  Our main motivation for
the present work (and for our next one) was to propose a general
approach to syntax (and ultimately to semantics) accounting for this
example in the spirit of our previous work
\cite{Hirschowitz-Maggesi-2010}. This is achieved thanks to our second
main result (Theorem \ref{t:arity-lim-colim}) which states the
existence of colimits in the category of (strengthened) arities.

In this extended abstract, we do not discuss proofs. A complete
version is available on-line.%
\footnote{\href{http://web.math.unifi.it/users/maggesi/strengthened/}{http://web.math.unifi.it/users/maggesi/strengthened/}.}

\section{Related and future work}
\label{ss:related}

The idea that the notion of monad is suited for modeling substitution
concerning syntax (and semantics) has been retained by many recent
contributions on the subject (see e.g.~\cite{BPfold,GU06,MU03})
although some other settings have been considered.  For instance in
\cite{HP07} the authors argue in favor of a setting based on Lawvere
theories, while in \cite{FPT} the authors work within a setting
roughly based on operads (although they do not write this word down).
The latter approach has been broadly extended, notably by M.~Fiore
\cite {F08,DBLP:conf/csl/FioreH10,DBLP:conf/mfcs/FioreM10}.  Our main
specificity here is the systematic use of the observation that the
natural transformations we deal with are linear with respect to
natural structures of module (a form of linearity had already been
observed, in the operadic setting, see \cite{FT01}, Section 4).

The signatures we consider here are much more general than the
signatures in \cite{FPT}, and cover the signatures appearing in
\cite{MU03,GU06}.  Note however that the latter works treat also
non-wellfounded syntax, an aspect which we do not consider at all.

In our next work, we will propose a treatment of equational semantics
for the present syntaxes. This approach should also be accommodated to
deal with typed languages as done for elementary signatures in
\cite{Zsido, ju_phd, ahrens_zsido_JFR}, or to model operational
semantics as done for elementary signatures in
\cite{2011arXiv1107.5252A}.

\section{The big category of modules}
\label{s:modules}

Modules over monads and the associated notion of linear natural
transformation intend to capture the notion of ``algebraic structure
which is well-behaved with respect to substitution''.  An introduction
on this subject can be found in our papers
\cite{HM,Hirschowitz-Maggesi-2010}.  Let us recall here the very basic
idea.

Let $R$ be a monad over a base category $\CC$.  A module over $R$ with
range in a category $\DD$ is a functor $M\colon \CC\to \DD$ endowed
with an action of $R$, i.e., a natural ``substitution'' transformation
$\rho\colon M\cdot R \rar M$ compatible with the substitution of $R$
in the obvious sense.  Given two modules $M,N$ over the same monad and
with the same range, a linear natural transformation $\phi\colon M \to
N$ is a natural transformation of functors which is compatible with
the actions in the obvious sense.  This gives a category $\Mod^\DD(R)$
of modules with fixed base $R$ and range $\DD$.

It is useful for the present paper to consider a larger category which
collects modules over different monads.  For the following definition,
we fix a range category $\DD$.

\begin{definition}[The big module category]
  We define the big module category $\BMod^\DD_\CC$ as follows:
  \squishlist 
\item its objects are pairs $(R, M)$ of a monad $R$ on $\CC$ and an
  $R$-module $M$ with range in $\DD$.
\item a morphism from $(R, M)$ to $(S, N)$ is a pair $(f, m)$ where
  $f\colon R \rar S$ is a morphism of monads, and $m\colon M \rar
  f^*N$ is a morphism of $R$-modules (here $f^*N$ is the functor $N$
  equipped with the obvious structure of $R$-module).
  \squishend 
\end{definition}

\section{The category of arities}
\label{s:higher-order}

In this section, we give our new notion of arity. The destiny of an
arity is to have representations in monads.  A representation of an
arity $a$ in a monad $R$ should be a morphism between two modules
$\dom(a, R)$ and $\codom(a, R)$.  For instance, in the case of the
arity $a$ of a binary operation, we have $\dom(a, R)\coleq R^2$ and
$\codom(a, R)\coleq R$. Hence an arity should consist of two halves,
each of which assigns to each monad $R$ a module over $R$ in a
functorial way. However, in all our natural examples, we have
$\codom(a, R)= R$ as above.  Although this will no longer be the case
in the typed case (which we do not consider here), we choose to
restrict our attention to arities of this kind, where $\codom(a, R)$
is $R$.


From now on we will consider only monads over the category $\Set$ and
modules with range $\Set$.  For technical reasons, see Section
\ref{s:stren-arities}, we restrict our attention to the category of
$\omega$-cocontinuous endofunctors that we will denote
$\End^\omega(\Set)$. Analogously we will write $\Mon^\omega$
(resp. $\BMod^\omega$) for the full subcategory of monads (resp. of
modules over these monads) which are $\omega$-cocontinuous.

We recall that finite limits commute with filtered colimits in
$\Set$.  It follows that $\End^\omega(\Set)$ has finite limits and
arbitrary (small) colimits.  This is the key ingredient in the proofs
of $\omega$-cocontinuity for most of our functors.

\begin{definition}[Arities]
  An \emph{arity} is a right-inverse functor to the forgetful functor
  from the category $\BMod^\omega$ to the category $\Mon^\omega$.
\end{definition}

Now we give our basic examples of arities:
\squishlist
\item Every monad $R$ is itself a $R$-module.  The assignment $R
  \mapsto R$ gives an arity which we denote by $\Theta$.
\item
  The assignment $R \mapsto *_R$, where $*_R$ denotes the final module
  over $R$ is an arity which we denote by $*$.
\item
  Given two arities $a$ and $b$, the assignment $R \mapsto a(R)\times
  b(R)$ is an arity which we denote by $a \times b$ .  In particular
  $\Theta^2=\Theta\times\Theta$ is the arity of any (first-order) binary
  operation and, in general $\Theta^n$ is the arity of $n$-ary
  operations.
\item Given an endofunctor $F$ of $\Set$, we consider the
  \emph{derived} functor given by $F'\colon X\mapsto F(X+*)$.  It can
  be checked how when $F$ is a module so is $F'$.  Given an arity $a$,
  the assignment $R \mapsto a(R)'$ is an arity which we denote $a'$
  and is called \emph{derivative} of $a$.
\item Derivation can be iterated.  We denote by $a^{(n)}$ the $n$-th
  derivative of $a$.  Hence, in particular, we have $a^{(0)} = a$,
  $a^{(1)}=a'$, $a^{(2)} = a''$.
\item For each sequence of non-negative integers $s=(s_1,\dots,s_n)$,
  the assignment $R \mapsto R^{(s_1)}\times \cdots \times R^{(s_n)}$
  is an arity which we denote by $\Theta^{(s)}$.  Arities of the form
  $\Theta^{(s)}$ are said \emph{algebraic}.  These algebraic arities
  are those which appear in \cite{FPT}.
\item
  Given two arities $a$, $b$ their composition $a\cdot b := R \mapsto
  a(R)\cdot b(R)$ is an arity.
\squishend


\begin{definition}
  \label{d:arity-morph}
  A morphism among two arities $a_1, a_2 \colon \Mon^\omega \rar \BMod^\omega$
  is a natural transformation $m \colon a_1 \rar a_2$ which, post-composed
  with the projection $\BMod^\omega \rar \Mon^\omega$, becomes the
  identity.  We easily check that arities form a subcategory $\Ar$ of the
  category of functors from $\Mon^\omega$ to $\BMod^\omega$.
\end{definition}

Now we give two examples of morphisms of arities:
\squishlist
\item
  The natural transformation $\mu \colon \Theta \cdot \Theta \rar
  \Theta$ induced by the structural composition of monads is a
  morphism of arities.
\item
  The two natural transformations $\Theta \cdot \eta$ and $\eta \cdot
  \Theta$ from $\Theta$ to $\Theta \cdot \Theta$ are morphisms of
  arities.
\squishend



\begin{theorem}
  \label{t:arity-lim-colim}
  The category of arities has finite limits and arbitrary (small)
  colimits.
\end{theorem}

\section{Categories of representations}
\label{s:representations}


\begin{definition}[Signatures]
  \label{d:signatures}
  We define a signature $\Sigma =(O, \alpha)$ to be a family of
  arities $\alpha \colon O \rar \Ar$.  A signature is said to be
  algebraic if it consists of algebraic arities.
\end{definition}

\begin{definition}[Representation of an arity, of a signature]
  \label{d:representations}
  Given an $\omega$-cocontinuous monad $R$ over $\Set$, we define a
  representation of the arity $a$ in $R$ to be a module morphism from
  $a(R)$ to $ R$; a representation of a signature $\Sigma$ in $R$
  consists of a representation in $R$ for each arity in $\Sigma$.
\end{definition}

\begin{example}
  The usual $\app\colon \Lambda^2 \rar \Lambda$ is a representation of
  the arity $\Theta^2$ into the monad $\Lambda$ of $\lambda$-calculus
  \ref{s:examples}.
\end{example}


\begin{definition}
  Given a signature $\Sigma = (O, \alpha)$, we build the category
  $\Mon^{\Sigma}$ of representations of $\Sigma$ as follows.  Its
  objects are $\omega$-cocontinuous monads equipped with a
  representation of $\Sigma$.  A morphism $m$ from $(M, r)$ to $(N,
  s)$ is a morphism of monads from $M$ to $N$ compatible with the
  representations in the sense that, for each $o$ in $O$, the
  following diagram of $M$-modules commutes:
  \[
    \xymatrix{
      \alpha_o(M) \ar[r]^{r_o}\ar[d]_{a_o(m)} & M \ar[d]^m \\
      m^* (\alpha_o(N)) \ar[r]_{m^*{s_o}} & m^* N}
  \]
  where the horizontal arrows come from the representations and the
  left vertical arrow comes from the functoriality of arities and
  $m\colon M \rar m^* N$ is the morphism of monad seen as morphism of
  $M$-modules.
\end{definition}

These morphisms, together with the obvious composition, turn
$\Mon^\Sigma$ into a category which comes equipped with a forgetful
functor to the category of monads.

\label{ss:representability}

We are primarily interested in the existence of an initial object in
this category $\Mon^\Sigma$.

\begin{definition}
  \label{d:representable}
  A signature $\Sigma$ is said representable if the category
  $\Mon^\Sigma$ has an initial object, which we denote $\hat\Sigma$.
\end{definition}

\begin{theorem}
  \label{t:sig-representable}
  Algebraic signatures are representable.
\end{theorem}

For more details we refer to our paper \cite{HM} (Theorems 1 and
2).  We give below a more general result (Theorem \ref{t:initsynt}).

\section{Modularity and the big category of representations}
\label{ss:modularity}

It has been stressed in \cite{GU06} that the standard approach (via
algebras) to higher-order syntax lacks modularity.  In the present
section we show in which sense our approach via modules enjoys
modularity. The key for this modularity is what we call the big
category of representations.

Suppose that we have a signature $\Sigma = (O, a)$ and two
subsignatures $\Sigma_1$ and $\Sigma_2$ covering $\Sigma$ in the
obvious sense, and let $\Sigma_0$ be the intersection of $\Sigma_1$
and $\Sigma_2$.  Suppose that these four signatures are representable
(for instance because $\Sigma$ is algebraic or strengthened in the
sense defined below). Modularity would mean that the corresponding
diagram of monads
\[
  \xymatrix{
    \hat \Sigma_0\ar[r]\ar[d] & \hat\Sigma_1\ar[d]\\
    \hat \Sigma_2\ar[r] & \hat\Sigma}
\]
is a pushout.  The observation of \cite{GU06} is that this diagram of
raw monads is, in general, not a pushout.  Since we do not want to
change the monads, in order to claim for modularity, we will have to
consider a category of enhanced monads.  Here by enriched monad, we
mean a monad equipped with some additional structure, namely a
representation of some signature.

Our solution to this problem goes through the following ``big''
category of representations, which we denote by $\WRep$, where $R$ may
stand for representation or for rich:
\squishlist 
\item An object of $\WRep$ is a triple $(R, \Sigma, r)$ where $R$ is a
  monad, $\Sigma$ a signature, and $r$ is a representation of $\Sigma$
  in $R$.
\item A morphism in $\WRep$ from $(R_1, (O_1, a_1), r_1)$ to $(R_2,
  (O_2, a_2), r_2)$ consists of an injective map $i\coleq O_1 \rar
  O_2$ compatible with $a_1$ and $a_2$ and a morphism $m$ from $(R_1,
  r_1)$ to $(R_2, i^*(r_2))$, where $i^*(r_2)$ should be understood as
  the restriction of the representation $r_2$ to the subsignature
  $(O_1, a_1)$ where we pose $i^*(r_2)(o) := r_2(i(o))$.
\item It is easily checked that the obvious composition turns $\WRep$
  into a category.
\squishend 
Now for each signature $\Sigma$, we have an obvious functor from
$\Mon^\Sigma$ to $\WRep$, through which we may see $\hat \Sigma$ as an
object in $\WRep$.  Furthermore, an injection $i\colon \Sigma_1 \rar
\Sigma_2$ obviously yields a morphism $i_*\coleq \hat \Sigma_1 \rar
\hat \Sigma_2$ in $\WRep$.  Hence our `pushout' square of
signatures as described above yields a square in $\WRep$. The proof of
the following statement is straightforward.

Modularity holds in $\WRep$, in the sense that given a `pushout'
square of representable signatures as described above, the associated
square in $\WRep$ is a pushout again.

As usual, we will denote by $\WRep^\omega$ the full subcategory of
$\WRep$ constituted by $\omega$-cocontinuous functors.  It is easy to
check that the previous statement is equally valid in $\WRep^\omega$.
Indeed, recall that, by our definition, the initial representation of
representable signatures lies in $\WRep^\omega$.

\section{Strengthening signatures}
\label{s:stren-arities}

Guided by the ideas of Matthes and Uustalu \cite{MU03} we introduce
in our framework the notion of \emph{strengthened arity}.
For a category $\CC$, let us denote by $\End_*^\omega(\CC)$ the
category of $\omega$-cocontinuous \emph{pointed endofunctors}, i.e.,
the category of pairs $(F, \eta)$ of an $\omega$-cocontinuous
endofunctor $F$ of $\CC$ and a natural transformation $\eta\colon
I\rar F$ from the identity endofunctor to $F$.  A morphism of pointed
endofunctors $f\colon (F_1,\eta_1) \rar (F_2,\eta_2)$ is a natural
transformation $f \colon F_1 \rar F_2$ satisfying $f\circ \eta_1 =
\eta_2$.

\begin{definition}
  A \emph{strengthened arity} is a pair $(H,\theta)$ where $H$ is an
  $\omega$-coconti\-nuous endofunctor of $\End^\omega(\Set)$ (i.e.,
  $H\in \End^\omega(\End^\omega(\Set))$) and $\theta$ is a natural
  transformation
  \(
    \theta \colon H(-)\cdot\sim \rar H(-\cdot\sim)
  \)
  (where $H(-)\cdot\sim$ and $H(-\cdot\sim)$ have to be understood as
  functors from $\End^\omega(\Set)\times\End_*^\omega(\Set)$ to
  $\End^\omega(\Set)$) satisfying
  \begin{math}
    \theta_{X,(I,1_I)} = 1_{HX}
 \end{math}
 and such that the following diagram is commutative
 \begin{equation}
   \label{e:theta_comp}
     \xymatrix@L=5pt{
       H(X)\cdot Z_1 \cdot Z_2
         \ar[rr]^{\theta_{X,(Z_1\cdot Z_2, e_1\cdot e_2)}}
         \ar[rd]_{\theta_{X,(Z_1,e_1)} Z_2} & &
       H(X\cdot Z_1\cdot Z_2) \\
       & H(X\cdot Z_1) \cdot Z_2 \ar[ru]_{\theta_{X\cdot Z_1,(Z_2,e_2)}}}
 \end{equation}
 for every endofunctor $X$ and pointed endofunctors $(Z_1,e_1)$,
 $(Z_2,e_2)$.
  We refer to $\theta$ as the \emph{strength} on $H$.
\end{definition}

Our first task is to make clear that our wording is consistent in the
sense that a strengthened arity $H$ somehow yields a genuine arity
$\tilde H$.  For this task, for each monad $R$ we pose $\tilde H (R)
\coleq H(R)$ and we exhibit on it a structure of $R$-module.  We do
even slightly more by upgrading $H$ into a \emph{module transformer}
in the following sense:
\begin{definition}
  \label{d:mod-trans}
  A module transformer is an endofunctor of the big module category
  $\BMod^\omega$ which commutes with the structural forgetful functor
  $\BMod^\omega \rar \Mon^\omega$.
\end{definition}

  Let $(H,\theta)$ be a strengthened arity.  For every
  $\omega$-cocontinuous monad $R$ and $\omega$-cocontinuous
  $R$-module $M$, we define the natural transformation
  $\rho^{H(M)}\colon H(M)\cdot R \rar H(M)$ as the composition
  $H(\rho^M) \cdot \theta_{M,R}$.
  Then $(H(M),\rho^{H(M)})$ is an $R$-module, and this construction
  upgrades $H$ into a module transformer denoted by $\hat H$.

We call the restricion $\tilde H$ of the module transformer $\hat H$
to the category of monads the arity associated to the strengthened
arity $H$.

Our next task is to upgrade our favorite examples of arities into
strengthened arities:
\squishlist
\item
  The arity $\Theta$ comes from the strengthened arity $(H, \theta)$
  where $H$ and $\theta$ are the relevant identities.
\item
  The arity $*$ comes from the strengthened arity $(H, \theta)$ where
  $H$ is the final endofunctor and $\theta$ is the relevant
  identity. This is the final strengthened arity.
\item
  The arity $\Theta \cdot \Theta$ comes from the strengthened arity
  $(H, \theta)$ where $H \coleq X \mapsto X \cdot X$ and $\theta_{X, Y}: X
  \cdot X\cdot Y \rar X\cdot Y \cdot X\cdot Y := X\cdot \eta^Y \cdot
  X\cdot Y$; here we have written $\eta^Y$ for the morphism from the
  identity functor to $Y$ (remember that $Y$ is pointed).
\item
  If an arity comes from a strengthened arity, so does its derivative
  (see Proposition \ref{p:stren-derived}).
\squishend


Then we show how our basic constructions in the category of arities
carries over the category of strengthened arities.  First we describe
this category.  Its objects are strengthened arities and we take for
morphisms from $(H_1, \theta_1)$ to $(H_2, \theta_2)$ those natural
transformations $m\colon H_1 \rar H_2$ which are compatible with
$\theta_1$ and $\theta_2$, that is, the diagram
\[
  \xymatrix@R=14pt{
    H_1(X)\cdot Z \ar[r]^{\theta_1}\ar[d]_{m_XZ} & H_1(X\cdot Z) \ar[d]^{m_{X\cdot Z}}\\
    H_2(X)\cdot Z \ar[r]_{\theta_2} & H_2(X\cdot Z)}
\]
is commutative for every endofunctor $X$ and every pointed endofunctor
$Z$.

\begin{theorem}
  \label{t:stren-lim-colim}
  The category of strengthened arities has finite limits and arbitrary
  colimits.
\end{theorem}

Next, we take care of the derivation.  We denote by $\DD$ the
endofunctor of $\Set$ given by $A\mapsto A+*$.  For any other
pointed endofunctor $X$ over $\Set$ we have a natural transformation
$w^X\colon D\cdot X \rar X \cdot D$ given by
\[
  w^X_A \colon X(A)+*\rar X(A+*)  \qquad
  w^X_A \coleq X(i_A)+\eta_{A+*}\cdot \underline *
\]
where $i_A\colon A\rar A+*$ and $\underline *\colon * \rar A+*$ are
the inclusion maps.

\begin{proposition}
  \label{p:stren-derived}
  If $(H,\theta)$ is a strengthened arity, then the pair
  $(H',\theta')$, where $H' \coleq X \mapsto H(X)'$ and
  $\theta'_{X,Z} \coleq \theta_{X,Z} D \cdot H(X) w^Z$, is a
  strengthened arity.  We call it the \emph{derivative} of
  $(H,\theta)$.
\end{proposition}

Now we point out the possibility of composing strengthened arities.

\begin{definition}
  \label{d:stren-comp}
  If $H :=(H,\rho)$ and $K :=(K,\sigma)$ are two strengthened arities,
  their composition $H \cdot K$ is the pair $(H\cdot K, \theta)$ where
  $\theta$ is defined by $\theta_{X,(Z,e)} \coleq H(\sigma_{X,(Z,e)})
  \cdot \rho_{K(X),(Z,e)}$.
\end{definition}

\begin{proposition}
  \label{p:stren-comp}
  This composition turns strengthened arities into a strict monoidal category.
\end{proposition}


Next, we turn to the main interest of strengthened arities (or
signatures) which is that the fixed point we are interested in
inherits a structure of monad.

\begin{lemma}
  Let $(H,\theta)$ be a strengthened arity.  Then the fixed point $T$
  of the functor $F \coleq I + H$ is $\omega$-cocontinuous and comes
  equipped with a structure of $\tilde H$-representation which is the initial
  object in the category of the $\tilde H$-representations.
\end{lemma}

We say that a signature is strengthened if it is a family of
strengthened arities.
The previous lemma leads immediately to the following result.
\begin{theorem}
  \label{t:initsynt}
  Strengthened signatures are representable.
\end{theorem}

\section{Examples of strengthened syntax}
\label{s:examples}

\paragraph{Lambda-calculus modulo $\alpha$-equivalence}
\label{ss:lambda-calc-alpha}

One paradigmatic example of syntax with binding is the
$\lambda$-calculus.  We denote by $\Lambda(X)$ the set of lambda-terms
up to $\alpha$-equivalence with free variables `indexed' by the set
$X$.  It is well-known \cite{BPdebruijn,Alt-Reus,HM} that $\Lambda$
has a natural structure of cocontinuous monad where the monad
composition is given by variable substitution.

It can be easily verified that application \( \app\colon \Lambda^2
\rar \Lambda \) and abstraction \( \abs\colon \Lambda' \rar \Lambda \)
are $\Lambda$-linear natural transformations, that is, $\Lambda$ is a
monad endowed with a representation $\rho$ of the signature $\Sigma =
\{\app\colon\Theta^2, \abs\colon\Theta'\}$.
The monad $\Lambda$ is initial in the category $\Mon^\Sigma$ of
$\omega$-cocontinuous monads equipped with a representation of the
signature $\Sigma$.

This is an example of algebraic signature and thus already treated by
other previous works \cite{HM,Hirschowitz-Maggesi-2010,FPT}.  Here we
simply remark that our new theory covers such a classical case.

\paragraph{Explicit composition operator}
\label{ss:ex-expl-comp}

We now consider our first example of non-algebraic signature.  On any
monad $R$, we have the composition operator (also called \textsf{join}
operator)
\(
  \mu^R \colon R \cdot R \rar R
\)
which has arity $\Theta \cdot \Theta$.  We will refer to the $\mu^R$
operator as the \emph{implicit} composition operator.  An interesting
problem is to see if this kind of operators admits a corresponding
\emph{explicit} version, i.e., if they can be implemented as a
syntactic construction.  As we have seen before 
$\Theta\cdot\Theta$ is a strengthened arity hence we can build
syntaxes with explicit composition operator of kind
\[
  \join\colon\Theta\cdot\Theta\rar\Theta.
\]

Of course, this is only a \emph{syntactic} composition operator, in
the sense that it does not enjoy several desirable conversion rules
like associativity, two-side identity and the obvious compatibility
rules with the other syntactic constructions present in the signature.
In our next work we will show how to construct such kind of
\emph{semantic} composition operator.

Let us mention that given a monad $R$, the unit $\eta_R\colon I \rar
R$ is not an $R$-linear morphism (in fact, $I$ is not even an
$R$-module in general).  For this reason we cannot treat examples of
syntax with explicit unit.

\paragraph{Syntax and semantics with explicit substitution}
\label{ss:ex-expl-subst}

On any monad $R$, we have a series of substitution operators
\(
  \sigma_n \colon R^{(n)} \cdot R^n \rar R
\)
which simultaneously replace $n$ formal arguments in a term with $n$
given terms.  As observed by Ghani and Uustalu \cite{GU06}, these
substitution morphisms satisfy a series of compatibility relations
which mean that they come from a single morphism \( \subst \colon C
\rar \Theta \) where $C$ is identified as the coend
\[
  C = \int^{A:\Fin} \Theta^{(A)}\times \Theta^A .
\]
Here $\Fin$ stands for a skeleton of the category of finite sets,
$\Theta^A$ denotes the cartesian power and $\Theta^{(A)}$ is defined
by $\Theta^{(A)}(R, X) \coleq R(X + A)$. Since coends are special
colimits, and strengthened arities admit colimits, we just have to
check that the bifunctorial arity $(A, B) \mapsto \Theta^{(A)}\times
\Theta^B$ factors through the category of strengthened arities.  As
far as objects are concerned, this follows from our results in Section
\ref{s:stren-arities}.  The verification of the compatibility of the
corresponding ``renaming'' and ``projection'' morphisms with the
strengthened structures is straightforward.

\bibliographystyle{eptcs}
\bibliography{hoat}

\end{document}